\documentclass[ 
    aps,
    twocolumn,
    groupedaddress,
    nofootinbib,
    longbibliography,
    balancelastpage
]{revtex4-2}

\usepackage[
    colorlinks=true,
    urlcolor=blue,
    anchorcolor=blue,
    citecolor=blue,
    filecolor=blue,
    linkcolor=blue,
    menucolor=blue,
    linktocpage=true,
    pdfproducer=medialab,
    pdfa=true
]{hyperref}

\usepackage{graphicx,color,amsmath,amssymb,flushend,bm,mathrsfs}

\usepackage[capitalise]{cleveref}
\crefname{section}{Sec.}{Secs.}

\usepackage{tikz}
\usepackage{ulem}
\usetikzlibrary{calc}
\usepackage{siunitx}
\DeclareSIUnit{\year}{yr}
\DeclareSIUnit{\parsec}{pc}

\newcommand{\refcite}[1]{Ref.~\cite{#1}}
\newcommand{\refscite}[1]{Refs.~\cite{#1}}

\newcommand{\bb}[1]{\bm{\mathrm{#1}}}
\newcommand{\du}{\mathrm{d}}
\newcommand{\dd}{\,\du}
\newcommand{\fobs}{f}
\newcommand{\fs}{f_{\mathrm{s}}}
\newcommand{\EGW}{E_{\mathrm{GW}}}
\newcommand{\AGW}{A_{\mathrm{GW}}}
\newcommand{\mmed}{m}
\newcommand{\mred}{\mu}
\newcommand{\specidx}{\beta}
\newcommand{\MG}[1]{M^{\mathrm G}_#1}

\begin{document}

\title{Discovering new forces with gravitational waves from supermassive black holes}

\author{Jeff A. Dror}
\email{jdror1@ucsc.edu}

\author{Benjamin V. Lehmann}
\email{benvlehmann@gmail.com}

\author{Hiren H. Patel}
\email{hpatel6@ucsc.edu}

\author{Stefano Profumo}
\email{profumo@ucsc.edu}

\affiliation{Department of Physics, University of California Santa Cruz, 1156 High St., Santa Cruz, CA 95064, USA
\\
and Santa Cruz Institute for Particle Physics, 1156 High St., Santa Cruz, CA 95064, USA}

\begin{abstract}
Supermassive black hole binary mergers generate a stochastic gravitational wave background detectable by pulsar timing arrays. While the amplitude of this background is subject to significant uncertainties, the frequency dependence is a robust prediction of general relativity. We show that the effects of new forces beyond the Standard Model can modify this prediction and introduce unique features into the spectral shape. In particular, we consider the possibility that black holes in binaries are charged under a new long-range force, and we find that pulsar timing arrays are capable of robustly detecting such forces. Supermassive black holes and their environments can acquire charge due to high-energy particle production or dark sector interactions, making the measurement of the spectral shape a powerful test of fundamental physics.
\end{abstract}

\maketitle

\section{Introduction}
\label{sec:introduction}
Pulsar timing arrays (PTAs) are on the verge of a historic discovery: the detection of a stochastic gravitational wave background (SGWB) produced by supermassive black hole (SMBH) binaries. PTAs use the extremely stable timing of successive light pulses from pulsars to detect gravitational waves (GWs) in the form of correlated timing distortions. In the presence of GWs, the observed time between pulses deviates from the stable rhythm in the frame of the source. The correlation of these deviations between pulsars exhibits a characteristic dependence on their angular separation, known as the Hellings and Downs curve~\cite{1983ApJ...265L..39H}, and this is considered the hallmark of a GW detection. 

Three major pulsar timing collaborations are currently searching for the GW background: NANOGrav~\cite{Arzoumanian:2018saf}, the European Pulsar Timing Array (EPTA)~\cite{Lentati:2015qwp}, and the Parkes Pulsar Timing Array (PPTA)~\cite{Shannon:2015ect}.
The sensitivities of these experiments vary based on their observational samples. Recently, the NANOGrav experiment has found evidence for a statistically significant correlated signal among a collection of $\mathcal O(50)$ pulsars in its 12.5-year dataset~\cite{Arzoumanian:2020vkk}. This may be the first signal of the SGWB from SMBH mergers, which would mark a monumental event in the history of GW astronomy. While the signal does not yet conclusively exhibit the Hellings and Downs angular dependence, upcoming datasets from NANOGrav and the other collaborations will be able to definitively confirm or refute the prospective discovery. Regardless of the fate of this particular signal, upcoming radio telescopes such FAST~\cite{2011IJMPD..20..989N} and SKA~\cite{Carilli:2004nx} will be sensitive to stochastic backgrounds well below even the most pessimistic predictions for the SGWB amplitude~\cite{Bonetti:2017lnj}, so a conclusive detection is expected in the near future.

Here, we point out that beyond astrophysical and cosmological applications, the study of the SGWB from SMBH mergers will open an entirely new observable for particle physics: the spectral shape of the SGWB.
If binaries are driven to merge by gravitational radiation alone, then the frequency dependence of the SGWB is cleanly predicted to be a power law with a known index. We show that physics beyond the Standard Model can modify this prediction of the spectral shape, and we thus evaluate the possibility of using forthcoming observations of the SGWB from SMBHs to test fundamental physics.

Previous work on using GW emission to detect new forces in binaries has been focused on pulsar systems~\cite{Krause:1994ar,Alexander:2018qzg,Dror:2019uea} such as the Hulse--Taylor binary~\cite{1975ApJ...195L..51H} and individual binaries with $\mathcal O(\SI{}{M_\odot})$ masses~\cite{Croon:2017zcu,Kopp:2018jom,Alexander:2018qzg,Fabbrichesi:2019ema,Dror:2019uea,Yue:2020bcf,Xu:2020qek} using recent detections made by the LIGO/Virgo Collaborations~\cite{TheLIGOScientific:2017qsa,Abbott:2020khf}. However, the unique evolution and environmental properties of SMBH binaries make them a rich laboratory to search for physics beyond the Standard Model. SMBHs copiously accrete nearby matter as they grow and they can develop powerful jets that accelerate particles to energies well above the electroweak scale~\cite{Romero:2020usl}. This may result in accretion or production of new particles, resulting in these SMBHs and their surroundings acquiring exotic quantum numbers. Moreover, SMBHs are known to be surrounded by a diffuse dark matter (DM) halo, and if this halo influences the dynamics of SMBH binaries, then these objects may be sensitive to dark matter interactions.

Such scenarios illustrate a range of new physics that may be discovered in SMBH mergers. Thus, the imminent measurement of the SGWB opens up an opportunity to access a wide range of physics inaccessible to terrestrial experiments. In this paper, we focus on the simple possibility that SMBHs in merging binaries carry a charge under a new (`fifth') force with a long but possibly finite range. We study how the standard power-law prediction for the SGWB spectrum is modified in the presence of such a new force, presenting the spectral index as a robust prediction.

For simplicity, we assume that the SMBHs themselves are charged, but our results also hold if a bound cloud of charge surrounds each SMBH. Our main assumption is that the charge distribution near each SMBH is pointlike on the scale of the binary separation. Detailed mechanisms for the accumulation of charge on and near SMBHs will be the subject of future work. We emphasize that any new physics which impacts the dynamics of merging binaries is potentially observable via the SGWB spectrum. We study a new long-range force as a benchmark scenario because this case is easily parametrized and demonstrates the key implications for the SGWB spectrum, but similar techniques can be applied in a variety of other scenarios.

This paper is organized as follows. In \cref{sec:binary}, we review the dynamics of a single binary in the presence of a new force. In \cref{sec:astrophysics} we present a calculation of the stochastic spectrum, highlighting various potential systematic uncertainties. We present our results in \cref{sec:discussion} in light of current constraints and the recent NANOGrav measurement.

Throughout this work, we denote the binary component masses by $M_1$ and $M_2$. We use $\omega$ for the orbital angular frequency of the binary, $\fs$ for the GW frequency in the frame of the source, and $\fobs$ for the observed GW frequency.

\section{The spectrum of SMBH mergers}
\label{sec:binary}
The SGWB from SMBH mergers has been studied extensively in the absence of new physics \cite{Mingarelli:2019pyd,Ryu:2018yhv,Bonetti:2017lnj,Enoki:2006kj,Sesana:2010qb,Sesana:2012ak,Sesana:2013wja}. Despite the complex astrophysical environments of SMBH mergers, the shape of the resulting SGWB spectrum can be predicted cleanly for a simple reason: the SGWB is dominated by contributions from binaries in the final stages of inspiral, where binary evolution is dominated by the emission of gravitational radiation with little pollution from environmental influences. Thus, there is a tight relationship between the radiated power and the hardening of the binary, leading in turn to a robust prediction for the shape of the GW spectrum.

The amplitude of the SGWB spectrum at a given frequency $f$ can be described in several ways. To facilitate comparison with existing literature, we discuss the spectrum in terms of the characteristic strain $h_c$. This is related to the energy density $\Omega_{\mathrm{GW}}$ by \cite{Moore:2014lga}
\begin{equation}
    \label{eq:energy-density-strain}
    h_c^2(f) = \frac{3H_0^2}{2\pi^2f^2}\Omega_{\mathrm{GW}}(f) \equiv \biggl[\AGW \times \Bigl(\frac{f}{\text{yr}^{-1}}\Bigr)^\specidx \biggr]^2
    \,,
\end{equation}
where $H_0$ is the Hubble parameter, $\AGW$ is the dimensionless amplitude (the value of the characteristic strain evaluated at an inverse-year), and $\specidx$ is the spectral index. For a spectrum that is not a power law, we allow $ \specidx $ to be frequency-dependent. 

The full SGWB spectrum can then be computed by combining the spectra of individual mergers over cosmic time. Following \refcite{Phinney:2001di}, the characteristic strain of the SGWB observed at a frequency $\fobs$ is given by
\begin{equation}
    \label{eq:sgwb-spectrum}
    h_c^2(\fobs) = \frac{3H_0^2}{2\pi^2\rho_c\fobs^2}
        \int\du z\dd\bb X\,\frac{\du n_{\mathrm{s}}}{\du z\dd\bb X}
        \frac{\fs}{1+z}
        \left.\frac{\du \EGW}{\du\fs}\right|_{\bb X}
    \,,
\end{equation}
where $n_{\mathrm{s}}$ is the comoving number density of GW sources, $\fs = (1+z)\fobs$ is the frequency in the frame of the source, and $\du \EGW/\du\fs$ is the energy spectrum produced by a single source. Here $\bb X$ denotes the state variables needed to determine the spectrum of a single source. If the sources are circular SMBH binaries, then in the absence of new physics, $\bb X$ simply denotes the component masses.\footnote{
    The binary separation is not an additional parameter of SMBH binary sources, since the spectrum $\du \EGW/\du\fs$ is obtained by integrating over all stages of binary evolution.
} As shown in the Appendix, the dominant contribution to the integral arises from redshifts of $ z \lesssim 0.3   $ and SMBH masses between $ \SI{e8}{M_\odot}  $ and $ \SI{e9}{M_\odot} $.

In the frequency range accessible to PTAs, the observable SGWB signal is expected to be dominated by SMBH binaries in the late stages of inspiral, where gravitational radiation is the primary mechanism for the binary to lose mechanical energy. As we review below, a merger driven by gravitational radiation alone produces a GW spectrum with shape $\du \EGW/\du\fs\propto \fs\null^{-1/3}$. This means that the frequency dependence in \cref{eq:sgwb-spectrum} can be factored out of the integral, and we obtain $h_c\propto \fobs^{-2/3}$. Thus $\specidx = -2/3$ independent of the properties of the binary population.

Our central result is that new forces between the binary components can modify this spectral shape by modifying the single-merger spectrum $\du \EGW/\du\fs$. We now turn to the calculation of this spectrum in the presence of new physics, starting with the calculation of the spectrum $\du\EGW/\du\omega$ as a function of the orbital frequency $\omega$.

The shape of the spectrum $\du\EGW/\du\omega$ is caused by the rise in orbital frequency $\omega$ as the SMBH separation $r$ falls over time, according to
\begin{equation}
    \label{eq:GW-spec-def}
    \frac{\du\EGW}{\du\omega} =
        \frac{\du\EGW}{\du t} \frac{\du t}{\du r} \frac{\du r}{\du \omega}
    \,.
\end{equation}
The orbital frequency $\omega$ is fixed by the SMBH separation through central forces acting between the binary components, via
\begin{equation}
    \label{eq:notKepler}
    \mred \omega^2 r = F(r)
    \,,
\end{equation}
where $\mred = M_1 M_2/(M_1+M_2)$ is the reduced mass of the system. 
Orbital decay then occurs as the mechanical energy $E_\text{mech} = \frac{1}{2}\mred r^2 \omega^2 + U(r)$ of the binary is lost to radiation, with energy per unit time $ P _{\mathrm{rad}} $. Conservation of energy gives
\begin{equation}
    \label{eq:energy-cons}
    0 = \frac{\du E_\text{mech}}{\du t} + P _{ \mathrm{rad}}
      = \mred r \omega^2\left(
            2+\frac{r}{\omega}\frac{\du \omega}{\du r}
        \right)\frac{\du r}{\du t} + P_\text{rad}
    \,.
\end{equation}
After solving \cref{eq:energy-cons} for $\du r/\du t$, substituting into \cref{eq:GW-spec-def}, and expressing the orbital frequency in terms of the radiation frequency as $\omega = \pi\fs$, the spectrum of GWs produced by the merger of a single binary is
\begin{equation}
    \label{eq:GW-spectrum}
    \frac{\du\EGW}{\du\fs} = - \pi^2 \mred r^2 \fs \left(
        \frac{2\fs}{r}\frac{\du r}{\du\fs} + 1
    \right)\frac{P_\text{GW}}{P_\text{rad}}
    \,,
\end{equation}
where $r$ is determined as a function of $\fs$ by \cref{eq:notKepler}.

In the case that the evolution of the binary is dominated by gravity, the central force is given by Newton's law, $F(r) = G M_1 M_2/r^2$, so that \cref{eq:notKepler} yields the well-known Kepler relation
\begin{equation} 
    \label{eq:Kepler}
    \omega ^2  = \frac{ G ( M _1 + M _2 ) }{  r ^3 }
    \,.
\end{equation} 
Meanwhile, orbital decay occurs predominantly through quadrupole radiation of gravity waves, with a power given by (see, e.g., Ref.~\cite{Maggiore:1900zz})
\begin{equation}
    P_\text{rad} = P_\text{GW} = \frac{32}{5} G \mred^2 \omega^6 r^4
    \,.
\end{equation}
The spectrum of GWs radiated in a single merger, \cref{eq:GW-spectrum}, becomes
\begin{equation}
    \label{eq:unmodified-single-spectrum}
    \frac{\du\EGW}{\du\fs} =
    \frac{\mred}{3}\Bigl[\frac{\pi^2 G^2(M_1 + M_2)^2}{\fs}\Bigr]^{1/3}
    \,,
\end{equation}
which exhibits a spectral index of $-1/3$. This gives rise to the spectral index of $\specidx=-2/3$ for the characteristic strain spectrum $h_c$ when integrated over cosmic time, according to \cref{eq:sgwb-spectrum}.

We now consider the effects of a new force mediated by a particle of mass $\mmed$ on the SGWB spectrum, similar to the treatment of individual neutron star binaries in \refcite{Croon:2017zcu}. We emphasize that our main assumption is that the charge distribution remains pointlike relative to the binary separation, which is of order \SI{e-2}{\parsec} in the PTA window. Thus, we do not require that the SMBHs themselves are charged. Still, we note that the particle nature of the additional species has important implications for charge stability in any concrete model where the SMBHs are directly charged. Firstly, charged black holes can neutralize by emission of charged particles. For Standard Model electric charge, this process is very slow and can be neglected for SMBHs with masses of order \SI{e9}{M_\odot} \cite{Gibbons:1975kk}. This may or may not be the case for the new charge as well, depending on the mass and coupling of the lightest charged state. Secondly, for a vector mediator with $\mmed>0$ or for a scalar mediator, no-hair theorems suggest that charge deposited directly onto an SMBH is not stable. For a massive vector, the effective charge of the SMBH decays on a timescale of order $\mmed^{-1}$ \cite{Coleman:1991ku}. We will be interested in extremely light mediators, corresponding to a relatively long timescale for decay. If the SMBHs are charged by a mechanism that remains active throughout the evolution of a binary, then no-hair theorems imply a reduction in the equilibrium charge, but do not necessarily preclude significant charges on the SMBHs themselves. On the other hand, a force mediated by a scalar can act directly on the SMBHs only in exceptional circumstances.

We now proceed to compute the GW spectrum in the presence of the new force, regardless of the particle nature of the interaction or whether the SMBHs are directly charged. The net force between the SMBH binary components is modified by the addition of a short-range contribution given by
\begin{equation}\label{eq:modifiedForce}
    F = \frac{G M_1 M_2}{r^2}\Bigl(1-\alpha e^{-\mmed r}(1+\mmed r)\Bigr)
    \,,
\end{equation}
where the potential-strength parameter $\alpha$ parametrizes the strength of the new force. We use the convention that the force is repulsive if $ \alpha > 0 $. The potential-strength parameter is given by
\begin{equation}
    \label{eq:potential-strength-parameter}
    \alpha = \frac{Q_1 Q_2}{G M_1 M_2}
    \,,
\end{equation}
where $Q_1$ and $Q_2$ are effective dark charges on the SMBHs. The normalization is chosen so that $\alpha=1$ for two extremal black holes if they were directly charged. Note, however, that the effective charges might arise from a charged cloud of particles surrounding the SMBHs. We defer a more detailed treatment of this scenario to future work.

The new force can supply another contribution to energy loss in the form of dipole radiation, $P_\text{dip}$.  The precise dependence on the frequency depends on the spin of the new mediator~\cite{Krause:1994ar},
\begin{equation}
    \label{eq:p-dipole}
    \textstyle
    P_\text{dip} = \frac{1}{3}G \gamma^2 \mu^2 r^2 \omega^4
    \sqrt{1-\frac{\mmed^2}{\omega^2}}\times
    \begin{cases}
        1-\frac{\mmed^2}{\omega^2}, & \text{(scalar)}
        \\[2mm]
        2+\frac{\mmed^2}{\omega^2}, & \text{(vector)}
    \end{cases}
\end{equation}
where the dimensionless dipole-strength parameter $\gamma$ characterizes the strength of radiation and is given in terms of the SMBH charges and masses by
\begin{equation}
    \label{eq:dipole-strength-parameter}
    \gamma ^2  = \frac{1}{G}\left(\frac{Q_1}{M_1} - \frac{Q_2}{M_2}\right)^2
    \,.
\end{equation}

Since nonzero $\gamma$ sources dipole radiation, its effect on energy loss is parametrically enhanced relative to the quadrupole gravitational radiation.  The enhancement is given by
\begin{equation}
    \frac{P_\text{dip}}{P_\text{GW}} =
    \frac{5 \gamma^2}{48 r^2 \omega^2} \simeq 
    20 ~\gamma^2 \Bigl(\frac{\SI{e9}{M_\odot}}{M_1+M_2}\Bigr)^{2/3}
        \Bigl(\frac{\SI{}{\year^{-1}}}{\omega}\Bigr)^{2/3}
    \,,
\end{equation}
where, in the second approximation, $r$ has been traded for $\omega$ using \cref{eq:Kepler}, assuming $\alpha=0$ and $\mmed=0$.  This shows that for $ \gamma = 1 $, the power lost to dipole radiation is about 20 times larger than that lost to gravitational radiation for the GW frequencies probed by pulsar timing experiments.

\section{New forces in the SGWB spectrum}
\label{sec:astrophysics}
We now predict the SGWB spectrum in the presence of new forces and compare the novel spectral features to astrophysical systematics.

We compute the observed strain using \cref{eq:sgwb-spectrum} in combination with \cref{eq:GW-spectrum}. This calculation requires an estimate of the number density of merging SMBH binaries, $n_{\mathrm s}$. SMBH binaries form when their host galaxies merge and their central BHs sink to the center by dynamical friction \cite{Begelman:1980vb}, so the SMBH merger rate depends on the galactic merger rate. The abundance and properties of galaxy pairs can be inferred from astronomical observations, and empirical scaling relations can then be used to connect galaxy properties to the properties of their resident SMBHs. To compute the SGWB spectrum including the normalization, we follow the procedure detailed in \refcite{Sesana:2012ak}, taking the galaxy mass function from \refcite{Ilbert:2009ub,Bell:2003cj}, the black hole--bulge mass relation from \refcite{McConnell:2012hz}, and the pair fraction from \refcite{LopezSanjuan:2012ea}. We give the full details of this calculation in the Appendix. Different choices of observational data from the literature produce variations in the normalization of the spectrum, resulting in a factor of $\mathcal O(10)$ uncertainty in the prediction of $\AGW$. However, again, these uncertainties pertain to the normalization of the SGWB, and not its spectral shape.

\subsection{Spectral features from new forces}

\begin{figure*}\centering
    \begin{tikzpicture} 
    \node(L) at (-4.5,0) {\includegraphics[width=8.5cm]{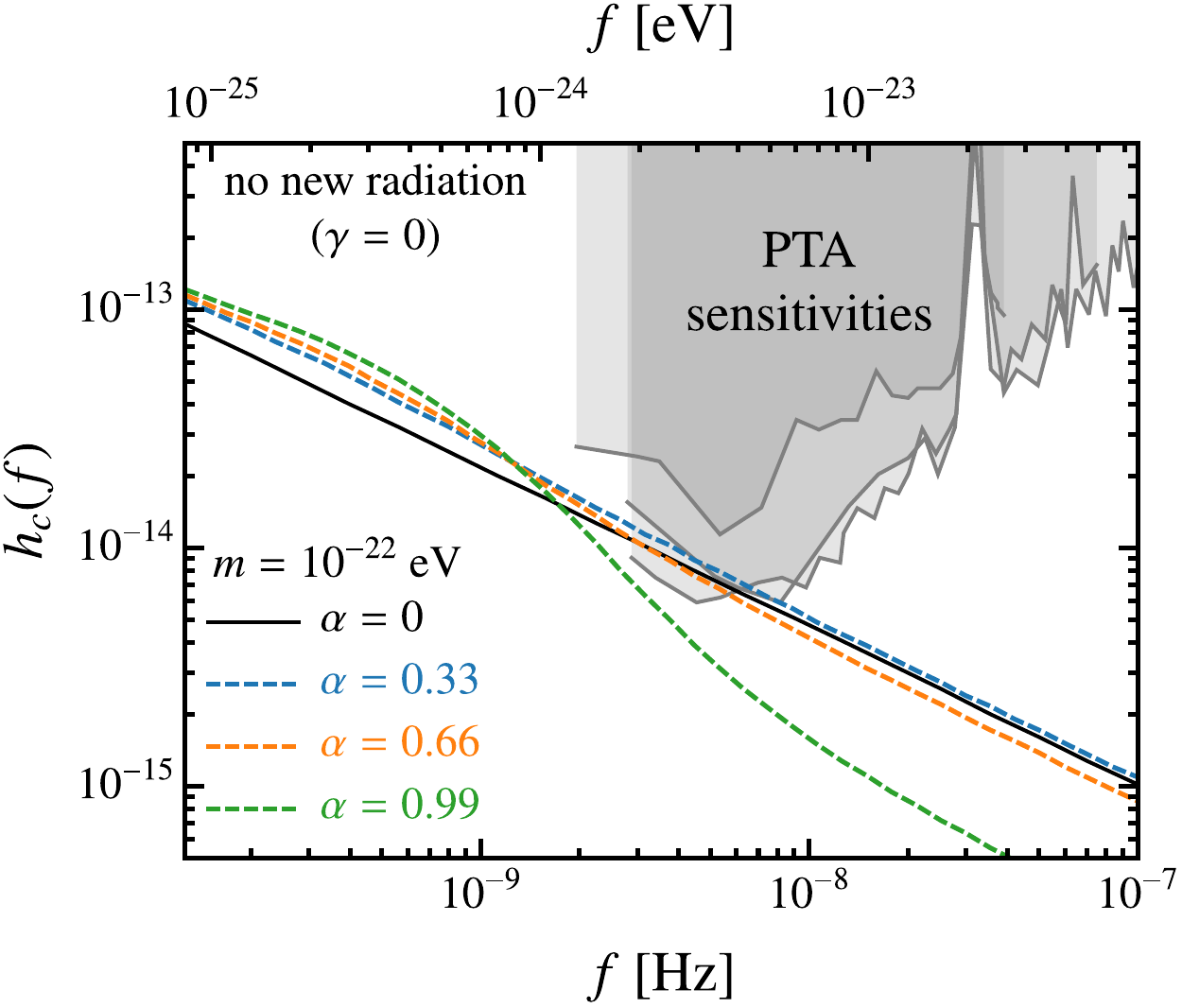}};
    \node(R) at (4.5,0){\includegraphics[width=8.5cm]{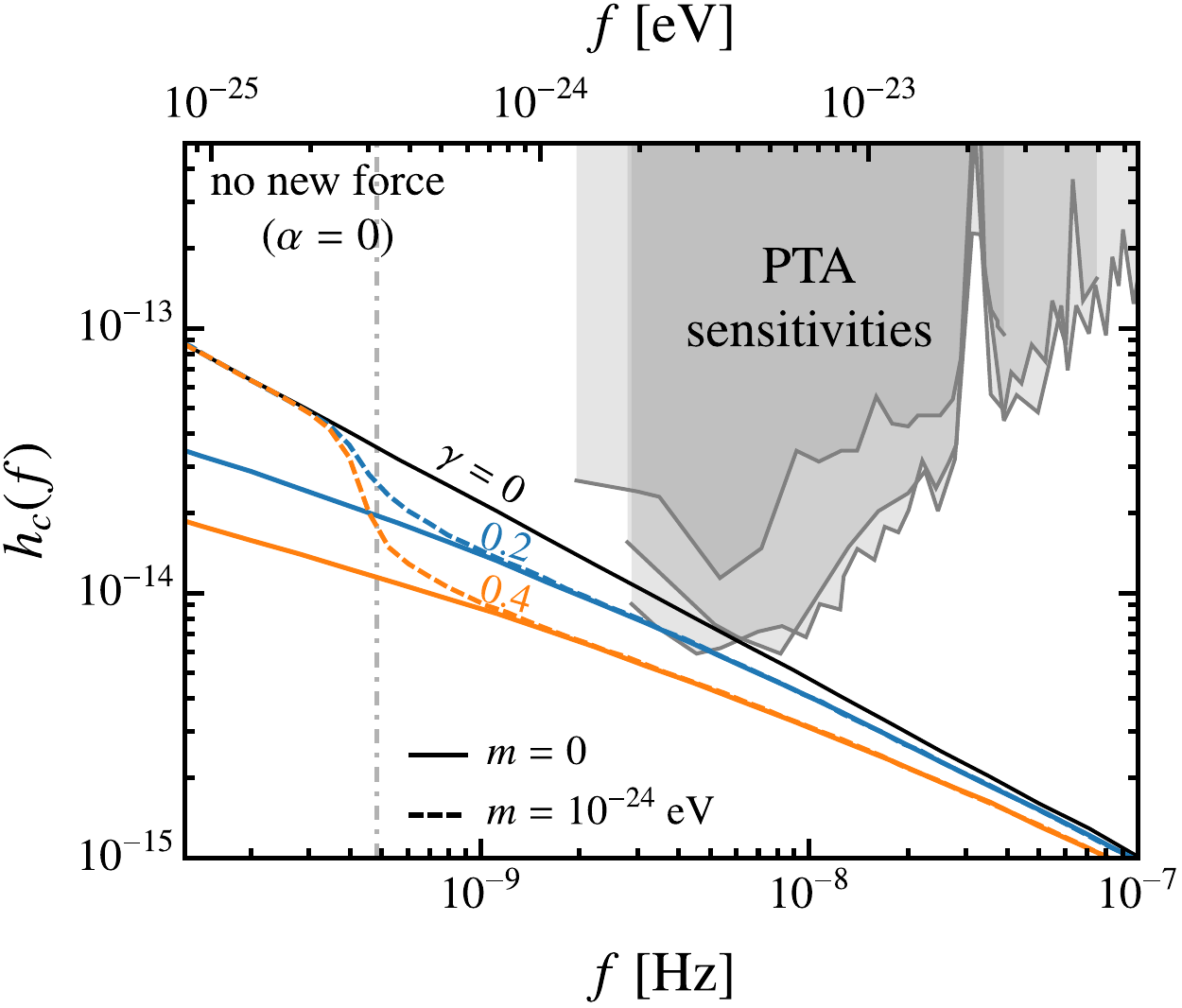}};
    \end{tikzpicture}
    \caption{Predicted SGWB produced by a population of uniformly charged SMBH binaries. The gray regions show current PTA sensitivities \cite{Arzoumanian:2018saf,Lentati:2015qwp,Shannon:2015ect}. In each panel, the black line shows the gravity-only prediction. {\bf Left:} SGWB produced with nonzero potential-strength parameter $ \alpha $, defined in \cref{eq:potential-strength-parameter}, for a mediator mass of $m=\SI{e-22}{\electronvolt}$. The location of the feature in the spectrum corresponds to black hole radial separations of order $m^{-1}$, and thus the location of the feature is offset from the mediator mass, as given by \cref{eq:feature-location-alpha}. {\bf Right:} SGWB produced with nonzero dipole-strength parameter $ \gamma $, defined in \cref{eq:dipole-strength-parameter}, for a vector mediator with mass of \SI{e-24}{\electronvolt}. The dot-dashed vertical line at $f_{\mathrm{th}} = m/\pi$ indicates the threshold for dipole radiation in the source frame.}
    \label{fig:OmegaVariation}
\end{figure*}

SGWB spectra with nonzero values of $\alpha$ and $\gamma$ are shown in the left and right panels of~\cref{fig:OmegaVariation}, respectively. In both cases, the spectral index of the SGWB is modified from the gravity-only prediction of $h_c\propto\fobs^{-2/3}$ shown in black, and non-power-law behavior may be directly observable in certain regimes. Note that we assume that all binaries have the same values of $\alpha$ and $\gamma$, but nontrivial distributions can be studied by taking $\bb X=\{M_1,M_2,Q_1,Q_2\}$ in \cref{eq:sgwb-spectrum}.  Below we discuss effects on the SGWB spectrum from nonzero $\alpha$ and nonzero $\gamma$ separately.
 
In the case with $\alpha\neq0$ and $\gamma=0$ (left panel of \cref{fig:OmegaVariation}), each SMBH carries the same nonzero dark charge. Thus, there is a new force between the two objects in addition to gravity. This modifies the usual form of Kepler's law relating the binary separation to its orbital frequency. For a massless mediator $\mmed=0$ and $\gamma=0$, the effect of the new force (\cref{eq:modifiedForce}) leads to a rescaling of Newton's gravitational force law by $G \rightarrow G(1-\alpha)$, and therefore preserves the shape of the SGWB spectrum while modifying only its normalization $\AGW$.  In this situation, it would be difficult to differentiate between new physics effects and astrophysical uncertainties in $\AGW$. However, for a mediator with nonzero mass $\mmed$, a distinctive feature emerges, as shown in \cref{fig:OmegaVariation}.  Since the nongravitational force is ineffective at separations $r > \mmed^{-1}$, the spectrum departs from a power law at a frequency corresponding to this separation, given by
\begin{align}    
    \nonumber
    f_* &= \sqrt{G (M_1 + M_2) m^3}/\pi +\mathcal{O}(\alpha)\\
    \label{eq:feature-location-alpha}
    &\simeq \SI{e-24}{\electronvolt} \left(\frac{M_1 + M_2}{\SI{e9}{M_\odot}}\right)^{1/2}\left(\frac{m}{\SI{e-22}{\electronvolt}}\right)^{3/2}
    .
\end{align}

In the case with $\alpha=0$ and $\gamma\neq0$ (right panel of \cref{fig:OmegaVariation}), only one of the two SMBHs is charged, so there is no modification to the force law which relates binary separation and orbital frequency. However, since the dark charge distribution now has a sizable dipole moment, the binary can lose energy to dipole radiation of the light mediator (here we assume a vector mediator such that we employ the lower case of \cref{eq:p-dipole}). The losses to dipole radiation can easily exceed those to gravitational radiation, which is sourced by the quadrupole moment. This leads to significant modification of the spectral index.

As with the $\alpha\neq0$ case, the $\gamma\neq0$ case exhibits additional features when the mediator mass $\mmed$ is nonzero. In this case, the spectrum reveals a threshold $\omega = m$ above which dipole radiation becomes significant. The binary rapidly loses energy above this threshold, producing a steplike feature in the spectrum around $\fobs\sim \mmed/\pi $. For a single merger, this feature is sharp, arising from the square root in \cref{eq:p-dipole}. It is slightly smoothed out in \cref{fig:OmegaVariation} by integration over the SMBH binary population across different redshifts.

If $\gamma\neq0$ and the mediator is massless, then the spectrum sourced by a single binary can be parametrized as
\begin{equation}
    \frac{\du\EGW}{\du\fs} \propto
    f^{-1/3} \frac{1}{1 + (f/\kappa)^{-2/3}}
    \,,
\end{equation}
where $\kappa$ is given by 
\begin{equation}
    \kappa \simeq \SI{2e-8}{\hertz}\left(\frac{\gamma}{0.2}\right)^3 \left(
        \frac{M_1+M_2}{\SI{e9}{M_\odot}}
    \right)^{-1}
    \,.
\end{equation}
Thus, for a given binary, the spectrum transitions between spectral indexes $\beta=-2/3$ and $\beta=-1/3$ at a frequency of order $\kappa$. This transitional behavior is also visible in the integrated SGWB shown in \cref{fig:OmegaVariation}, in which the curves with $\gamma > 0$ exhibit a shallower power-law index at frequencies $f\ll\SI{e8}{\hertz}$. 

In general, an SMBH binary can have both a nonzero $ \alpha $ and a nonzero $ \gamma $ such that the spectrum will be a combination of the two limiting cases presented here.

\subsection{Distinguishing new forces from astrophysics}
Predictions of the amplitude and shape of the SGWB spectrum from SMBH mergers are sensitive to astrophysical uncertainties associated with galactic mergers. Even in the gravity-only calculations, the details of the cosmological population of such mergers determines the normalization of the spectrum. We seek to probe spectral features which vary between different binaries at different redshifts. Thus, it is essential to assess the full SGWB spectrum obtained by convolving the single-merger spectrum of \cref{eq:GW-spectrum} with the statistics of galactic mergers.

The prediction of \cref{eq:sgwb-spectrum} depends on the validity of the single-merger spectrum. In the standard scenario, the single-merger spectrum of \cref{eq:GW-spectrum} is valid only for circular binaries that are driven to merge purely by emission of gravitational radiation. In the absence of new physics, these assumptions are robust in the late stages of inspiral, for which the SGWB is most prominent, and it is this very simplicity that makes the SGWB spectrum such a powerful probe of new forces. However, even in the absence of new forces, there are other astrophysical processes that modify parts of the SGWB spectrum. These modifications represent possible systematics, or, if they are well understood, they may provide a new set of features for extensions of our analysis. We thus briefly outline these processes and the implications for the SGWB spectrum. Most importantly, these processes mainly influence the spectrum of GWs outside the window probed by pulsar timing arrays.

Typically, the dynamics of a binary are governed by a single energy loss mechanism at any given time. At very large separations, corresponding to low frequencies in the SGWB, gravitational radiation is inefficient. The inspiral is instead driven by dynamical friction from stars and \cref{eq:unmodified-single-spectrum} is invalid: the binary shrinks much faster than would be expected from gravitational radiation alone, and thus the SGWB is suppressed at these frequencies. At small separations, stellar dynamical friction becomes inefficient as the binary depletes the region of stellar phase space that can remove energy from the binary (the ``loss cone'' \cite{Milosavljevic:2002bn}). At this point, in the absence of any other energy loss mechanism, gravitational radiation dominates the evolution of the binary for the remainder of the inspiral. The transition from stellar dynamical friction to gravitational radiation domination takes place at a characteristic separation corresponding to a frequency of order~\cite{Begelman:1980vb,Quinlan:1996vp}
\begin{equation}
    \label{eq:df-gw-transition}
    f_{\mathrm{GR}}
    \simeq\SI{e-9}{\hertz}
        \left(\frac{M_1M_2}{(\SI{5e8}{M_\odot})^2}\right)^{-3/8}
        \left(\frac{M_1+M_2}{\SI{e9}{M_\odot}}\right)^{1/8}
    \,,
\end{equation}
assuming that gas and stars shrink the binary on a timescale $\left|r/\dot r\right|\sim\SI{e8}{\year}$. The merger itself imposes an upper cutoff on the frequency corresponding to the separation of the binary at the innermost stable circular orbit (ISCO), given by 
\begin{equation}
    \label{eq:ISCO}
    f_{\mathrm{ISCO}} =  \frac{1}{2\pi}\frac{1}{6GM_1}
        \simeq \SI{11}{\micro\hertz}
            \left(\frac{M_1}{\SI{5e8}{M_\odot}}\right)^{-1}
    \,,
\end{equation}
where it is assumed that $M_1 = M_2$.

However, the spectrum of \cref{eq:unmodified-single-spectrum} is not typically valid at frequencies immediately above $f_{\mathrm{GR}}$ for two reasons: stalled mergers and eccentric orbits. At distances of order a parsec, energy loss from gravitational radiation is not efficient enough to merge a binary within the lifetime of the Universe. As a result, the evolution of merging binaries from the end of star-driven dynamical friction until the era where gravitational radiation becomes an efficient energy loss mechanism (separations below $\sim\SI{0.01}{\parsec}$) is not known. This is known as the ``final parsec problem''~\cite{Milosavljevic:2002bn}. Candidate mechanisms include gas dynamics~\cite{SanchezSalcedo:2000hd,Escala:2003dy,Escala:2004jh,Armitage:2002uu,Mayer:2007vk,Sesana:2012ak,Mayer:2013jja,Sesana:2013wja,Ravi:2014aha,Ravi:2014nua,Khan:2016vln,Burke-Spolaor:2018bvk} and asymmetry of galactic mergers~\cite{Berczik:2006tz}. In particular, efficient gas infall can dominate the evolution of the binary up to a frequency of several times $f_{\mathrm{GR}}$~\cite{Begelman:1980vb}. In the absence of any other mechanisms, binaries may even stall until a subsequent galactic merger supplies a third SMBH, at which point few-body dynamics can shrink the binary \cite{Bonetti:2017lnj,Ryu:2018yhv}. Such processes have a significant effect on the normalization of the SGWB, and gas dynamical processes may have a slight impact on the spectral shape as well at the lowest frequencies in the pulsar timing window~\cite{Taylor:2016ftv,Kelley:2016gse}. (See also \refscite{Ivanov:1998qk,Armitage:2002uu,Escala:2003dy,Escala:2004jh,Haiman:2009te,Mayer:2013jja,Aly:2015vqa} for further discussion of the role of gas dynamics in SMBH mergers.)

A more significant modification potentially arises from eccentricity of the orbits. The spectrum of \cref{eq:unmodified-single-spectrum} holds only for a circular binary. However, during the stage of inspiral driven by stellar dynamical friction, stellar encounters tend to enhance the eccentricity of the binary, so typical binaries in simulations have substantial eccentricities at $f_{\mathrm{GR}}$ \cite{Quinlan:1996vp}. These eccentricities are quickly reduced by gravitational radiation through a process termed ``circularization''~\cite{Enoki:2006kj,Sesana:2010qb}. Nevertheless, there is still a range of separations [and frequencies as given by \cref{eq:Kepler}] in which binaries are driven by gravitational radiation and yet deviating from \cref{eq:unmodified-single-spectrum}. This may change the spectral index of the SGWB in a range of frequencies above $f_{\mathrm{GR}}$. This spectral feature may extend to frequencies of order \SI{e-9}{\hertz}, or even as high as \SI{e-8}{\hertz} in some projections. In principle, this may mimic the effects of new physics. Claiming a discovery of a new force may require restricting analysis to GW frequencies above \SI{e-8}{\hertz}.

Finally, we note that realistic predictions of the SGWB are influenced by Poisson noise at frequencies above \SI{e-7}{\hertz}, as the SGWB is expected to be dominated by relatively few sources in this regime \cite{Taylor:2016ftv}. We neglect this effect in our analysis, as a modification to the spectral shape will still produce a significant modification to the spectrum of a small number of sources. However, a full statistical treatment in this regime should be performed using a Monte Carlo simulation rather than by direct measurement of the spectral index.

\section{Discussion}
\label{sec:discussion}
We have argued that the spectral index of the SGWB can be robustly predicted in the absence of new physics and have explored how a new force can modify the spectral index. We now discuss the implications of our results in light of the recent observation of a stochastic process among the pulsars in the NANOGrav 12.5-year dataset~\cite{Arzoumanian:2020vkk} as well as other GW detection experiments.

\begin{figure}\centering
    \includegraphics[width=0.9\columnwidth]{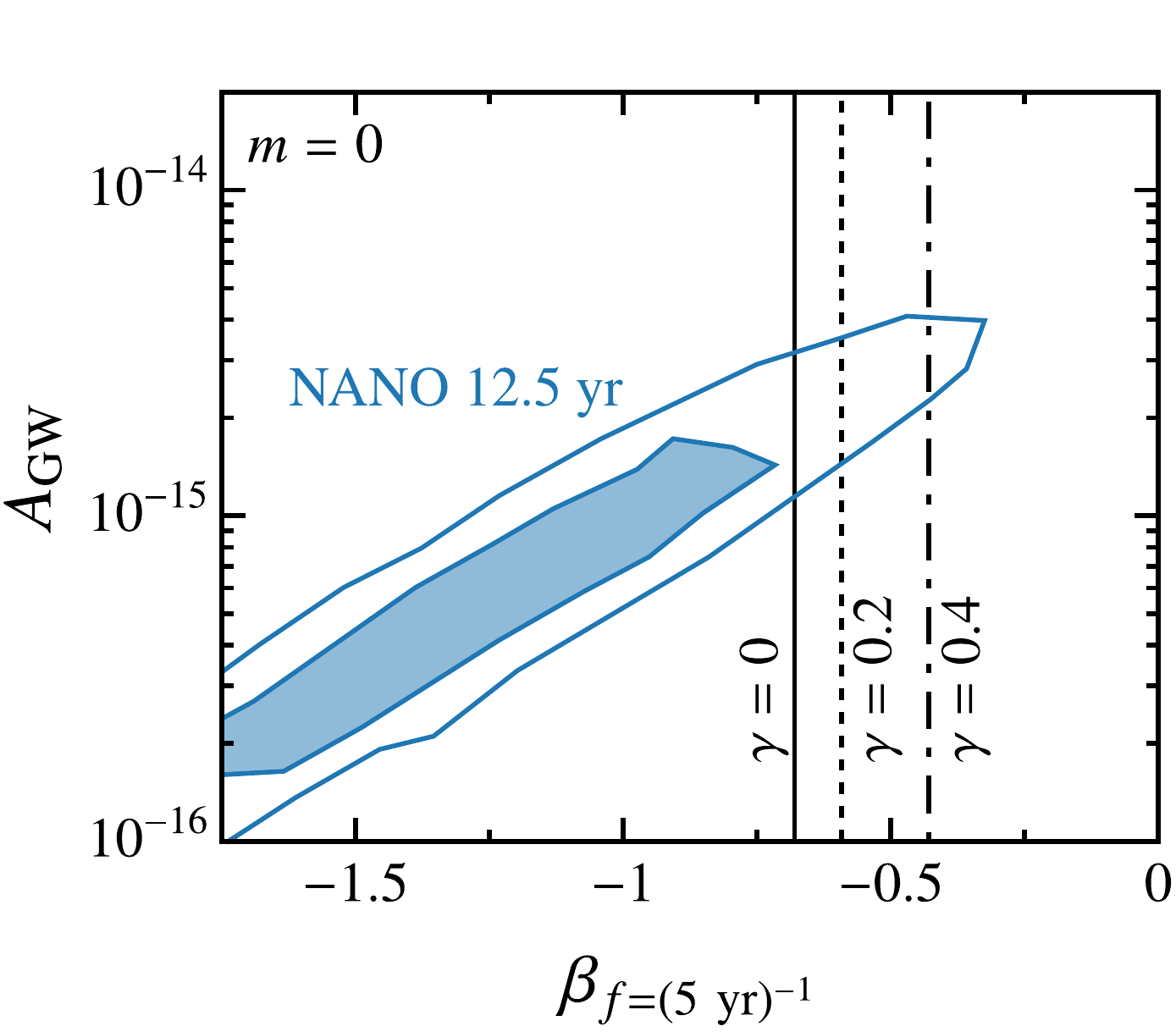}
    \caption{A comparison of the spectral index as measured in the NANOGrav 12.5-year data set to the value predicted by merging supermassive charged black hole binaries. The shaded and bounded regions correspond to the $1\sigma$ and $2\sigma$ posteriors derived by the NANOGrav Collaboration \cite{Arzoumanian:2020vkk}. The black lines correspond to charged binaries under a new long-range vector force with different values of the dipole-strength parameter $ \gamma $, assuming the potential-strength parameter is negligible ($ \alpha =  0 $). Since this spectrum is not strictly a power law, we evaluate the spectrum at roughly the peak sensitivity of NANOGrav, $ f =  (\SI{5}{\year})  ^{-1} $.}
    \label{fig:NANO}
\end{figure}

The NANOGrav Collaboration fits the spectrum to two types of power laws: one present in only the five lowest frequencies ($ \SI{2e-9}{\hertz}\lesssim \fobs \lesssim \SI{1e-8}{\hertz} $) and one present among the thirty lowest frequencies ($ \SI{2e-9}{\hertz}\lesssim \fobs \lesssim \SI{7e-8}{\hertz} $).~\footnote{The analysis is also carried out with a broken power law whose results are qualitatively similar to those of using just the five lowest frequencies.} While SMBH mergers are expected to produce GWs across the entire pulsar timing frequency band, pulsar-intrinsic noise may contribute at high frequencies and mask the GW signal~\cite{Arzoumanian:2020vkk}, and, as such, we focus on the five frequency analysis. The measured amplitude and spectral index for a power-law signal are depicted in \cref{fig:NANO}. The solid black line shows the predicted index for uncharged supermassive black holes. The spectrum of charged black holes is not generally a power law and can have various shapes depending on the mediator mass $ m $, the potential-strength parameter $ \alpha $, and the dipole-strength parameter $ \gamma $. In the limit where the mediator mass vanishes and the SMBHs carry a nonzero $ \gamma $, the spectrum is approximately a power law across the pulsar-timing window and we show the index evaluated at a frequency of $ ( \SI{5}{\year} ) ^{-1} $ in \cref{fig:NANO}. We conclude that the additional dipole radiation will soften the spectrum, and the current dataset can potentially constrain $ \gamma \sim 1 $.

While the uncertainties on the values of the amplitude and spectral index are still significant, pulsar timing arrays are rapidly improving in sensitivity. For identical pulsars, the signal-to-background ratio of a pulsar timing array analysis scales $ \propto \AGW ^2 T ^{ 13/3} N _p $, where $ T $ is the observation time and $ N _p $ is the number of pulsars~\cite{Moore:2014eua}. The large scaling with observation time suggest that NANOGrav will be able to significantly improve the estimate of the spectral index and amplitude as it continues observing the current pulsar set. Furthermore, combining the 12.5-year NANOGrav data with the EPTA and PPTA datasets may be enough to detect the Hellings and Downs correlation function between pulsars, which, if observed, would confirm the first detection of a stochastic GW background. Once a discovery is made, the measurement of the spectral index will be critical to measure the charges of the SMBHs and search for additional forces.

Pulsar timing arrays are particularly well suited to measure stochastic GW spectra at frequencies of order \SI{}{\nano\hertz}--\SI{100}{\micro\hertz}. SMBHs that are emitting GWs in this frequency band are near the start of their merger. As they progress toward the inspiral phase, the emission continues, building a falling characteristic strain spectrum until the ISCO frequency of the heaviest black holes, $\sim$\SI{e-5}{\hertz} [see~\cref{eq:ISCO}]. The measurement of the spectrum might be extendable using space-based interferometers such as the Laser Interferometer Space Antenna (LISA)~\cite{Seoane:2013qna,2017arXiv170200786A} or astrometry~\cite{1990NCimB.105.1141B}. Confirmation of a consistent spectral index and amplitude across this wide range of frequencies would be a remarkable confirmation of gravity-only mergers. On the other hand, if a new force is present with a mediator mass above the pulsar timing range and below that of higher frequency detectors, it would show up as an observable break in the spectrum. This displays the critical complementarity between the different GW searches.

Motivated by the imminent discovery prospects of a stochastic background of GWs in pulsar timing arrays, we have focused our discussion on the detection of new forces in SMBH binaries. However, other GW experiments may also detect stochastic binary merger backgrounds. In particular, LISA is expected to see a stochastic background of white dwarf, neutron star, and lighter black hole binary mergers~\cite{1987ApJ...323..129E,Bender_1997}. While these backgrounds are highly anisotropic, it is also possible to look for new forces in these new environments by incorporating directionality in \cref{eq:sgwb-spectrum}. The white dwarf background (as well as other stochastic merger backgrounds observed in the future) will provide complementary searches for dark forces in different astrophysical environments.

Finally, we note that since the GW spectrum from SMBH binaries is yet to be discovered, it is possible that SMBHs have charges so large that new force is strong relative to gravity. In this case, we may uncover additional signals in the SGWB. First, for sufficiently large dark charges, a repulsive force will stall the merger on cosmological timescales. This could reduce the SGWB amplitude below lower bounds estimated for gravity-only mergers~\cite{Bonetti:2017lnj}. Second, while gravitational radiation tends to rapidly circularize binaries, dipole radiation can have the opposite effect as the binary passes through the mediator mass threshold and can have a dramatic effect on the spectrum. Such phenomena do require a mechanism for the accumulation of large charges near or on the SMBHs. We leave the study of such mechanisms and their consequences for future work.

\section*{Acknowledgments}
We thank Enrico Ramirez-Ruiz and Luke Zoltan Kelley for useful discussions. J.A.D. is supported in part by NSF CAREER grant No. PHY-1915852. B.V.L., H.H.P., and S.P. are supported in part by DOE grant No. DE-SC0010107.

\appendix*

\section{Distribution of SMBH binaries}
\label{sec:astro-appendix}
The prediction of the SGWB spectrum relies on the statistical properties of the sources, which must be extracted from astronomical observations. This enters into the calculation of \cref{eq:sgwb-spectrum} via the quantity
\begin{equation}
    \label{eq:dns}
    \frac{\du n_{\mathrm s}}{\du z\dd\bb X} = 
    \frac{\du n_{\mathrm s}}{\du z\dd M_1\dd M_2}
    \,,
\end{equation}
where $n_{\mathrm s}$ is the comoving number density of SMBH binaries. The parameter distributions of such binaries are not measured directly. Instead, observations measure the population statistics of galaxies and the rate at which these galaxies merge. Following \refcite{Sesana:2012ak}, we combine these data with observational relations between galaxies and SMBHs to infer the properties of the SMBH binary population.

This process can be carried out in many different ways, using different astronomical datasets and SMBH--host relations. We now detail the method we use to predict the observed SGWB. As detailed in \refcite{Sesana:2012ak}, the various prescriptions introduce an order-of-magnitude uncertainty in the amplitude of the SGWB. The amplitude is relatively unimportant for the present study since we are most interested in modifications to the spectral shape. However, note that the spectral shape is in principle sensitive to the redshift distribution of sources, so uncertainties in the relative populations of sources at each redshift propagate to (small) uncertainties in the shape.

We infer the distribution of SMBH binary mergers from the distribution of galaxy mergers, under the assumption that each galaxy hosts an SMBH with mass related to the galaxy mass. Given an SMBH mass $M_i$, we will denote the host galaxy mass by $\MG{i}$. To better conform with the astronomy literature, we let $M_1$ denote the mass of the heavier SMBH, and we use the mass ratio $q\equiv \MG{2}/\MG{1} < 1$ instead of $\MG{2}$ directly. We write the differential merger rate of galaxies per unit comoving volume as 
\begin{equation}
    \label{eq:dng}
    \frac{\du n_{\mathrm G}}{\du z\dd\MG{1}\dd q} =
        \frac{\phi(\MG{1},z)}{\MG{1}\log10}
        \frac{1}{\tau(z, \MG{1}, q)}
        \frac{\du\mathscr F(z, \MG{1}, q)}{\du q}
        \frac{\du t}{\du z}
    \,.
\end{equation}
Here $\du t/\du z$ converts the rate with respect to time into a rate with respect to redshift; $\phi(M_{\mathrm G}) = \du n_{\mathrm G}/\du\log_{10} M_{\mathrm G}$ is the observed galaxy mass function as reported in the astronomy literature; $\tau(z,\MG{1},q)$ is the merger timescale of a given galactic pair; and $\mathscr F(z, M_{\mathrm G}, q)$ is the differential pair fraction, i.e., the fraction of galaxies of mass $M_{\mathrm G}$ in binaries with mass ratio $q$ at redshift $z$. We take the pair fraction directly from \refcite{LopezSanjuan:2012ea}. We take the galaxy mass function for $z>0.2$ from \refcite{Ilbert:2009ub}, and we interpolate from $z=0.2$ to the $z=0$ mass function of \refcite{Bell:2003cj}. We take the merger timescale $\tau$ from Eq.~(10) of \refcite{Kitzbichler:2008zj}, and following \refcite{Sesana:2012ak}, we include an additional factor of $\frac12q^{-0.3}$. We assume that the resulting SMBH binary merges instantaneously, so that \cref{eq:dns} and \cref{eq:dng} are equivalent up to the change of variable $(M_1,M_2)\mapsto(\MG{1},q)$. In other words, we compute the characteristic strain as 
\begin{multline}
    \label{eq:detail-spectrum}
    h_c^2(\fobs) = 
        \int\du z\dd \MG{1}\dd q\,
        \frac{\du n_{\mathrm G}}{\du z\dd \MG{1}\dd q}
        \frac{\fs}{1+z}
        \left.\frac{\du \EGW}{\du\fs}\right|_{\MG{1},q}
        \\\times
        \frac{3H_0^2}{2\pi^2\rho_c\fobs^2}
    .
\end{multline}

\begin{figure}\centering
    \includegraphics[width=\columnwidth]{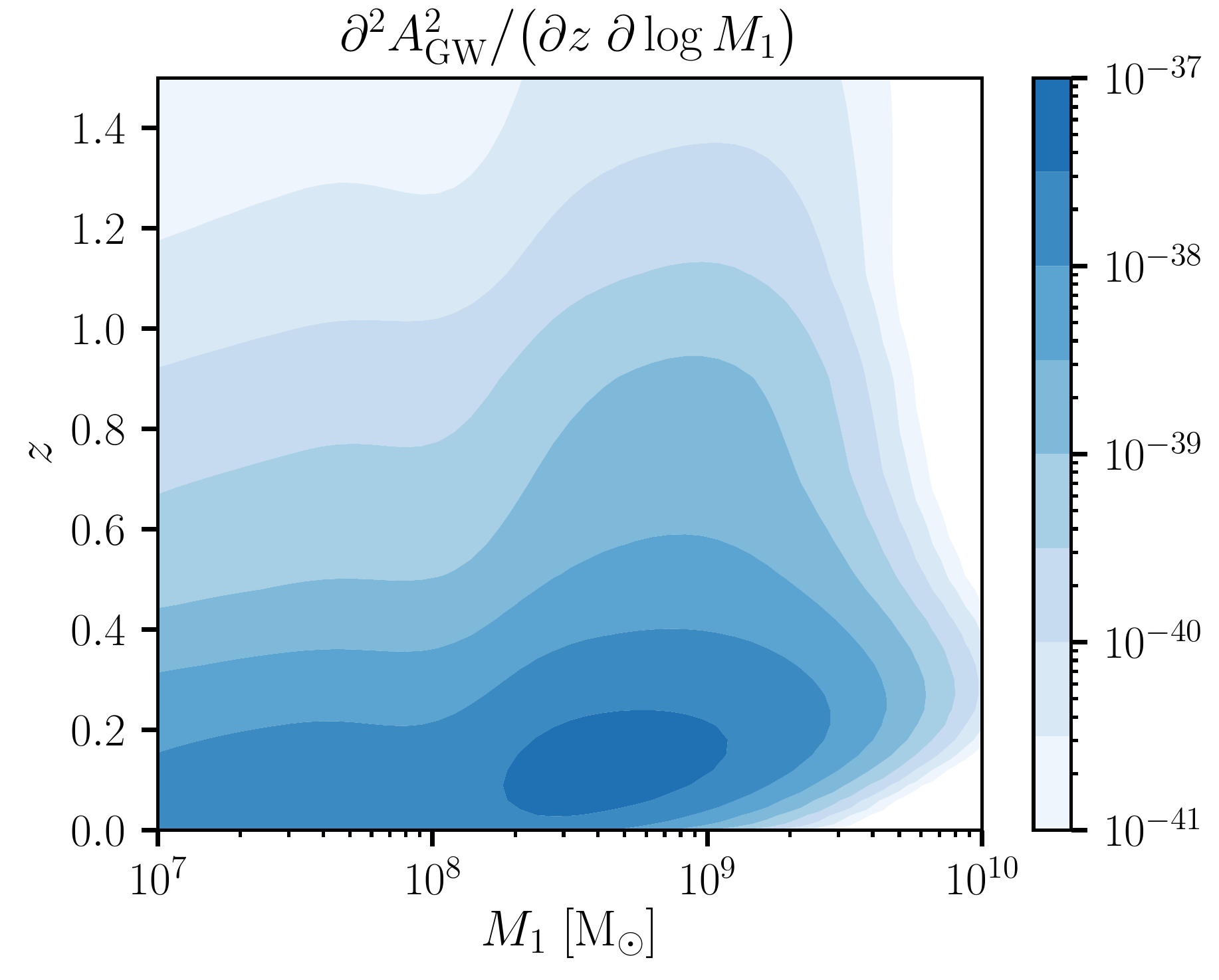}
    \caption{Differential contribution to the squared amplitude of the SGWB as a function of $M_1$ and $z$ in the gravity-only case. Here $M_1$ denotes the more massive component of the binary.}
    \label{fig:differential-contribution}
\end{figure}

To relate the galaxy mass to the SMBH mass, there are two steps: first, we estimate the stellar mass of the galactic bulge, $M_{\mathrm{bulge}}$. Second, we use the scaling relation of \refcite{McConnell:2012hz} to determine the SMBH mass from the bulge mass. We approximate the bulge fraction $f_{\mathrm{bulge}} \equiv M_{\mathrm{bulge}} / M_{\mathrm G}$ as a function of $M_{\mathrm G}$, following \refcite{Sesana:2012ak}: the bulge fraction should be $\sim0.9$ for galaxy masses above \SI{e11}{M_\odot}, and $\sim0.25$ for galaxy masses below \SI{e10}{M_\odot}. We smoothly interpolate between these two bulge fractions as $f_{\mathrm{bulge}}(M_{\mathrm G}) = 0.55 - 0.22\tan^{-1}(4 - 1.3\times10^{-10}M_{\mathrm G}/\SI{}{M_\odot})$.

This approach is useful for making a simple estimate with appropriate redshift and mass ratio dependence. However, certain choices must be made to arrive at a single prediction of the SGWB amplitude: \refcite{Sesana:2012ak} computes many projected strains based on different observed mass functions, pair fractions, and scaling relations. These estimates span more than a decade in $\AGW$, but the values are slightly lower than other estimates, and are used as a ``pessimistic'' projection by the NANOGrav Collaboration \cite{Arzoumanian:2018saf}. Therefore, when following the calculation of \refcite{Sesana:2012ak}, we make an optimistic set of choices to bring the normalization of the SGWB into line with the scenario considered to be ``moderate'' by the NANOGrav Collaboration. In particular, as discussed in that work, we add the uncertainties to the fitted mass function parameters and further add 0.1 dex (i.e., multiply by $10^{0.1}$) to account for systematics. We likewise add the scatter in the SMBH--host scaling relations to the fitted parameter values. We determine the masses of the SMBHs from the properties of the merged galaxy using the ``double accretion'' prescription of \refcite{Sesana:2008xk}.

\Cref{fig:differential-contribution} shows the differential contribution to the squared amplitude $\AGW^2$ of the SGWB as a function of $M_1$ and $z$ in the gravity-only case, i.e., the integrand of \cref{eq:detail-spectrum} evaluated at $\fobs = \SI{1}{\per\year}$. This indicates the relative contribution of different masses and redshifts to the SGWB signal given the observational data and scaling prescriptions used in this work. In particular, \cref{fig:differential-contribution} demonstrates that the signal is dominated by binaries with primary masses between \SI{e8}{M_\odot} and \SI{e9}{M_\odot} and redshifts $z \lesssim 0.3$.

\bibliographystyle{JHEP}
\bibliography{references}

\end{document}